\documentclass[preprint2,epsf,epsfig,graphics]{emulateapj}
\usepackage[breaklinks,colorlinks, urlcolor=blue,citecolor=blue,linkcolor=blue]{hyperref}
\usepackage{graphicx}
\usepackage{color}
\usepackage{amsmath}
\usepackage{color}

\shorttitle{Polarized foregrounds for EoR experiments II}
\shortauthors{Nunhokee et al.}

\begin{document}
\title{Constraining polarized foregrounds for EoR experiments II: polarization leakage simulations in the avoidance scheme}

\author{C.D.~Nunhokee\altaffilmark{1}}
\author{G.~Bernardi\altaffilmark{1,2}}
\author{S.A.~Kohn\altaffilmark{3}}
\author{J.E.~Aguirre\altaffilmark{3}}
\author{N.~Thyagarajan\altaffilmark{4}}
\author{J.S.~Dillon\altaffilmark{5,6,7}}
\author{G.~Foster\altaffilmark{1}}
\author{T.L.~Grobler\altaffilmark{1}}
\author{J.Z.E.~Martinot\altaffilmark{3}}
\author{A.R.~Parsons\altaffilmark{6,7}}

\email{cnunhokee@gmail.com}

\altaffiltext{1}{Department of Physics and Electronics, Rhodes University, PO Box 94, Grahamstown, 6140, South Africa}
\altaffiltext{2}{SKA SA, 3rd Floor, The Park, Park Road, Pinelands, 7405, South Africa}
\altaffiltext{3}{Department of Physics and Astronomy, University of Pennsylvania, Philadelphia, PA, USA}
\altaffiltext{4}{Arizona State University, School of Earth and Space Exploration, Tempe, AZ 85287, USA}
\altaffiltext{5}{Berkeley Center for Cosmological Physics, Berkeley, CA 94720, USA}
\altaffiltext{6}{Dept. of Astronomy, University of California, Berkeley, CA 94720, USA}
\altaffiltext{7}{Radio Astronomy Lab., U. California, Berkeley CA 94720, USA}

\begin{abstract}
A critical challenge in the observation of the redshifted 21-cm line is its separation from bright Galactic and extragalactic foregrounds. In particular, the instrumental leakage of polarized foregrounds, which undergo significant Faraday rotation as they propagate through the interstellar medium, may harmfully contaminate the 21-cm power spectrum. We develop a formalism to describe the leakage due to instrumental widefield effects in visibility--based power spectra measured with redundant arrays, extending the delay--spectrum approach presented in \cite{Parsons2012}. We construct polarized sky models and propagate them through the instrument model to simulate realistic full--sky observations with the Precision Array to Probe the Epoch of Reionization. We find that the leakage due to a population of polarized point sources is expected to be higher than diffuse Galactic polarization at any $k$ mode for a 30~m reference baseline. For the same reference baseline, a foreground--free window at $k > 0.3 \, h$~Mpc$^{-1}$ can be defined in terms of leakage from diffuse Galactic polarization even under the most pessimistic assumptions. If measurements of polarized foreground power spectra or a model of polarized foregrounds are given, our method is able to predict the polarization leakage in actual 21-cm observations, potentially enabling its statistical subtraction from the measured 21-cm power spectrum.
\end{abstract}

\keywords{cosmology: observations -- dark ages, reionization, first stars -- polarization -- techniques: interferometric}

% Necessary to get footnotes to number properly
\setcounter{footnote}{0}

\section{Introduction}
\label{sec:intro}
The study of the first luminous sources and the consequent epoch of reionization (EoR) occupies a central place in modern cosmology. Amongst the various probes, the redshifted 21-cm line is expected to be the most promising one, potentially illuminating our Universe even before the first stars started to shine \citep[see ][for recent reviews]{Furlanetto2016,McQuinn2016}. 

Observations of the redshifted 21-cm line have progressed tremendously over the last few years, with improving upper limits being placed over a wide redshift range from both sky--averaged \cite[][]{Bowman2010,Bernardi2016} and fluctuations \cite[][]{Dillon2015,Jacobs2015,Trott2016,Ewall-Wice2016, Beardsley2016,Patil2017} 21-cm observations. The most stringent upper limits on fluctuations to date are the power spectrum measurements at $z = 8.4$ from observations with the Precision Array to Probe the Epoch of Reionization \citep[PAPER,][]{Parsons2010} that provide evidence for heating of the Intergalactic Medium prior to reionization \cite[][]{Parsons2014,Ali2015,Pober2015,Greig2016}. 

The biggest challenge in measuring the 21-cm signal is the presence of foregrounds few orders of magnitude stronger than the cosmological signal \citep[e.g.][]{Jelic2008,Bernardi2009,Jelic2010,Zaroubi2012,Pober2013,Dillon2014,Parsons2014,Chapman2015}. A very effective technique in separating smooth--spectrum foregrounds and the 21-cm signal is the so called {\it foreground avoidance}: smooth--spectrum foregrounds occupy a wedge--shaped limited region of the two--dimensional power spectrum space whereas the remaining area -- the so--called EoR window -- is dominated by the 21-cm emission \citep[e.g.][]{Datta2010,Morales2012,Thyagarajan2013,Liu2014a,Liu2014b,Chapman2016}.

Deviations from frequency smoothness can, however spread power into the EoR window, potentially invalidating the avoidance assumptions. One such deviation has been recognized as leakage from Faraday rotated, polarized Galactic and extragalactic synchrotron emission \citep[][]{Bernardi2010,Jelic2010,Geil2011}. In particular, when the 21-cm power spectrum is estimated through the co-addition of redundant visibilities \citep[e.g.][]{Moore2013,Parsons2014}, leakage from polarized foregrounds due to asymmetric widefield primary polarization response cannot be corrected using similar approaches to interferometric imaging \citep[][]{Bhatnagar2008,Morales2009,Ord2010,Tasse2013,Asad2015} and it may ultimately jeopardize the 21-cm signal in the EoR window.

This paper follows the measurement of two dimensional polarized power spectra from \citet{Kohn2016} using PAPER observations. We extend the delay transform approach \citep{Parsons2012} used to avoid foregrounds in visibility--based 21-cm power spectra to the polarized case, defining a formalism that naturally includes the leakage due to widefield polarized primary beams. We develop models of polarized foregrounds to which we apply the formalism with the aim of simulating PAPER--like observations. We compared our simulated polarized power spectra with existing data and predict the leakage expected in 21-cm power spectra.

The paper is organized as follows: the formalism is derived in Section~\ref{sec:stokes}, simulations of actual observations are described in Section~\ref{sec:simulations}, and results discussed in Section~\ref{sec:results}. We conclude in Section~\ref{sec:conclusion}.

\section{Formalism}
\label{sec:stokes}
The relationship between the sky brightness distribution $s$ and the visibility ${\rm v}$ measured by a two--element interferometer is described by \citep[e.g.][]{thompson2008}:
\begin{equation}\label{eq:visibility_1}
{\rm v}({\boldsymbol{b}}, \nu)= \int_\Omega a(\hat{\boldsymbol{r}},\nu) \,
s(\hat{\boldsymbol{r}},\nu) \, e^{-2\pi i \nu \frac{{\boldsymbol{b}} \cdot \hat{\boldsymbol{r}}}{c} }\, d\Omega,
\end{equation}
where $s$ is the Stokes I parameter, ${\boldsymbol{b}} = (u,v,w)$\footnote{Throughout this paper, capital boldface letters are used to indicate matrices, small boldface letters to indicate vectors and small normal letters to indicate scalars.} is the vector representing the coordinates in meters in the plane of the array, $\hat{\boldsymbol{r}} = (l,m,n)$ is the unit vector representing the direction cosines on the celestial sphere \citep[see][for further details]{thompson2008}, $\nu$ is the observing frequency, $a$ describes the telescope primary beam response, $c$ is the speed of the light and $\Omega$ is the solid angle subtended by the source.
Throughout the paper we will assume array coplanarity and simulate zenith observations, leading to ${\boldsymbol{b}} = (u,v)$, $\hat{\boldsymbol{r}} = (l,m)$, $wn = 0$, and reducing equation~\ref{eq:visibility_1} to a two dimensional Fourier transform relationship.

Equation~\ref{eq:visibility_1} does not specify a polarization frame but it can be generalized to include the polarization state of the sky brightness using the measurement equation formalism \citep{Hamaker1996,Smirnov2011a}, where the input radiation field is related to the measured visibilities via $2 \times 2$ Jones matrices. If the intrinsic sky brightness distribution towards a line of sight $\hat{\boldsymbol{r}}$ at the frequency $\nu$ is described by the usual Stokes parameters ${\bf s}=(I,Q,U,V)^T$, with $T$ the transpose operator, the sky brightness distribution observed through the telescope primary beam ${\bf s'}=(I',Q',U',V')^T$ becomes \citep[e.g.][]{Ord2010}:
\begin{eqnarray}\label{eq:intfresp}
{\bf s}' (\hat{\boldsymbol{r}}, \nu)			& = & {\bf A}(\hat{\boldsymbol{r}}, \nu) \, {\bf s} (\hat{\boldsymbol{r}}, \nu) \nonumber \\
				& = & {\bf S}^{-1} \big[ {\bf J}(\hat{\boldsymbol{r}}, \nu) \otimes {\bf J}^* (\hat{\boldsymbol{r}}, \nu) \big] {\bf S} \, {\bf s} (\hat{\boldsymbol{r}}, \nu)
\end{eqnarray}
where ${\bf{J}}$ is the $2 \times 2$ Jones matrix representing the polarized primary beam response, $\otimes$ is the outer product operator, $^*$ denotes the complex conjugate and ${\bf{S}}$ is the matrix that relates the Stokes parameters to the orthogonal $x-y$ linear feed frame:
\begin{equation*}
{\bf{S}}=\frac{1}{2}\begin{pmatrix} 1 & 1 & 0 & 0 \\
0 &  0 & 1 &  i \\
0 &  0 & 1 & -i \\
1 & -1 & 0 &  0 \end{pmatrix}. 
\end{equation*}
The $4 \times 4$ matrix ${\bf A}$ can be seen as a mixing matrix between intrinsic and observed (primed) Stokes parameters:
\begin{equation*}
\begin{pmatrix}
I^{\prime} \leftarrow I & I^{\prime} \leftarrow Q & I^{\prime} \leftarrow U &  I^{\prime} \leftarrow V\\
Q^{\prime} \leftarrow I & Q^{\prime} \leftarrow Q & Q^{\prime} \leftarrow U &  Q^{\prime} \leftarrow V \\
U^{\prime} \leftarrow I & U^{\prime} \leftarrow Q & U^{\prime} \leftarrow U &  U^{\prime} \leftarrow V \\
V^{\prime} \leftarrow I & V^{\prime} \leftarrow Q & V^{\prime} \leftarrow U &  V^{\prime} \leftarrow V \\
\end{pmatrix}.
\end{equation*}
Equation~\ref{eq:visibility_1} can be extended to the polarized case by defining the four ``Stokes" visibility products ${\bf{v}} = ({\rm v}_{I},{\rm v}_{Q},{\rm v}_{U},{\rm v}_{V})^T$:
\begin{eqnarray}\label{eq:visibility_2}
{\bf v} ({\boldsymbol{b}}, \nu)	& = &	{\bf S}^{-1} \, {\bf v}_c ({\boldsymbol{b}}, \nu) \nonumber \\
				& = & {\bf S}^{-1} \int_{\Omega} \big[ {\bf J}(\hat{\boldsymbol{r}}, \nu) \otimes {\bf J}^* (\hat{\boldsymbol{r}}, \nu) \big] {\bf S} \, {\bf s} (\hat{\boldsymbol{r}}, \nu) \, e^{-2\pi i \nu \frac{{\boldsymbol{b}}.\hat{\boldsymbol{r}}}{c} }\, d\Omega \nonumber \\
				& = & \int_{\Omega} {\bf A}(\hat{\boldsymbol{r}}, \nu) \, {\bf s} (\hat{\boldsymbol{r}}, \nu) \, e^{-2\pi i \nu
                  \frac{{\boldsymbol{b}}.\hat{\boldsymbol{r}}}{c} }\, d\Omega
\end{eqnarray}
where ${\bf v}_c = ({\rm v}_{xx},{\rm v}_{xy},{\rm v}_{yx},{\rm v}_{yy})^T$ is the four cross--polarization correlator output. 
Equation~\ref{eq:visibility_2} recasts the full sky formalism developed by \cite{Smirnov2011a} in the Mueller matrix form.

We are now in the position to define four polarization power spectra by applying the delay transform to equation~\ref{eq:visibility_2}. 
The delay--transform is the Fourier transform of a single visibility along frequency \citep[][]{Parsons2009,Parsons2012}:
\begin{equation}\label{eq:delay_1}
\tilde{\rm v}({\boldsymbol{b}}, \tau)= \int_B {\rm w}(\nu) \, {\rm v}({\boldsymbol{b}}, \nu) \, e^{-2\pi i \nu\tau}\,d\nu\, ,
\end{equation}
where $B$ is the observing bandwidth, ${\rm w}$ is the window function and $\tau$ represents the geometric delay between antenna pairs:
\begin{equation}\label{eq:delay_2}
\tau = \frac{{\boldsymbol{b}} \cdot \hat{\boldsymbol{r}}}{c}.
\end{equation}
The delay transform is related to the 21-cm power spectrum $p(k)$ as \citep[][]{Parsons2012,Thyagarajan2016}:
\begin{align}\label{eq:power_spectrum}
 p(k) &= p \left ( \sqrt{k_\perp^2 + k_\parallel^2} \right ) \nonumber\\
 &=\Big(\frac{\lambda^2}{2k_{\rm B}}\Big)^2 \Big(\frac{D^2\Delta D}{B} \Big) \frac{1}{q} \, |{\tilde{\rm v}}(|{\boldsymbol b}|,\tau)|^2,
\end{align}
with
\begin{eqnarray}
k_\perp     & = &\frac{2\pi \frac{|\boldsymbol{b}|}{\lambda}}{D}\\
k_\parallel & = &\eta\frac{2\pi f_{21}H_0E(z)}{c \, (1+z)^2}
\end{eqnarray}
where $\lambda$ is the observing wavelength, $k_B$ is the Boltzmann constant, $D$ is the
transverse comoving distance, $\Delta D$ is the comoving depth along the line of sight corresponding to the bandwidth $B$, $f_{21}$ is the 21-cm line rest frequency, $z$ is the redshift,  $H_0$ is the Hubble constant and $E(z) = \sqrt{\Omega_M(1+z)^3 +\Omega_k(1+z)^2 + \Omega_\Lambda}$. In this work, we use $H_0=100\, h$~km~s$^{-1}$, $\Omega_M=0.27$, $\Omega_k = 0$ and $\Omega_\Lambda = 0.73$.
The power spectrum normalization volume $q$  is \citep{Thyagarajan2016}:
\begin{equation}
q= \int_{\Omega} \int_B |a(\hat{\boldsymbol{r}},\nu) \, {\rm w}(\nu)|^2 \, d\Omega \, d\nu.
\end{equation}

The polarized case is obtained by substituting equation~\ref{eq:visibility_2} in the delay transform:
\begin{equation}\label{eq:delay_3}
{\tilde{\bf{v}}}({\boldsymbol{b}}, \tau) = \int_B {\rm w}(\nu) \, {\bf v} ({\boldsymbol{b}}, \nu) \, e^{-2\pi i \nu\tau} \, d\nu\,
\end{equation}
and by using the Hadamard (or element--wise) product $\circ$ in order to extend equation~\ref{eq:power_spectrum} to four power spectra ${\bf p} = \big(p_I,p_Q,p_U,p_V)^T$:
\begin{eqnarray}\label{eq:power_spectrum_2}
{\bf p}(k) 	= \Big(\frac{\lambda^2}{2k_{\rm B}}\Big)^2 \frac{D^2\Delta D}{B} {\bf Q}^{-1} \left \{ {\tilde{\bf{v}}}(|{\boldsymbol b}|,\tau) \circ {\tilde{\bf{v}}}(|{\boldsymbol b}|,\tau)^* \right \}.
\end{eqnarray}
As the off--diagonal elements of the $\bf A$ matrix are much smaller than the diagonal elements, the normalization matrix $\bf Q$ may be written as a diagonal matrix with diagonal elements:
\begin{equation}\label{eq:omegabw}
\textrm{diag}({\bf Q}) \approx \int_{\Omega} \int_B \textrm{diag} \left ( {\bf A} \, {\rm w} \right ) \left [ \textrm{diag} ({\bf A}\, {\rm w}) \right ]^* \, d\Omega\, d\nu.
\end{equation}
Dimensionless polarized power spectra ${\bf \Delta}^2 = (\Delta_I^2,\Delta_Q^2,\Delta_U^2,\Delta_V^2)$ can be defined in analogy with the scalar case:
\begin{equation}\label{eq:power_spectrum_3}
{\bf \Delta}^2 (k) = \frac{k^3}{2 \pi^2} \, {\bf p} (k).
\end{equation}
Equation~\ref{eq:power_spectrum_2} is one of the main results of our work: if the sky emission is unpolarized, $p_I(k)$ reduces to $p(k)$, but appropriately combines the two orthogonal polarizations and their relative primary beams to form the total intensity power spectrum estimator. The remaining terms represent the visibility--based polarization power spectra which are dominated by leaked total intensity foreground emission.

In the presence of a polarized sky, $p_I(k)$ naturally includes the leakage of polarized emission due to widefield primary beams. Equation~\ref{eq:power_spectrum_2} therefore generalizes the approach of \cite{Moore2013} to full polarization and provides a framework to simulate the expected leakage to the 21-cm power spectrum given both a polarized foreground and an instrument model.

\section{Simulations}
\label{sec:simulations}
In order to evaluate equation~\ref{eq:power_spectrum_2} we need three ingredients: a model of the PAPER dipole beam, an array configuration and a polarized sky model.

\begin{figure*}
\centering
\includegraphics[width=1.\columnwidth]{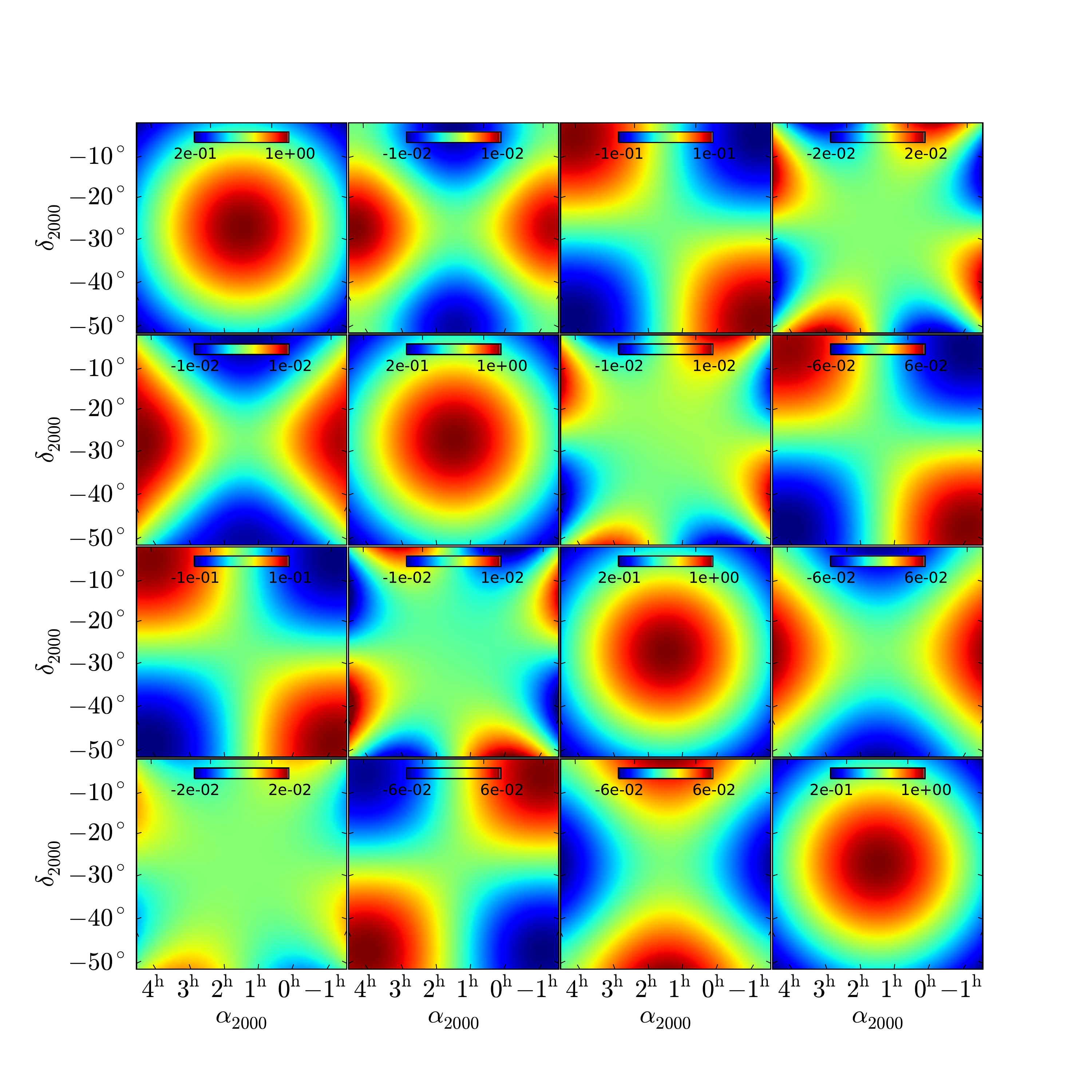}
\includegraphics[width=1.\columnwidth]{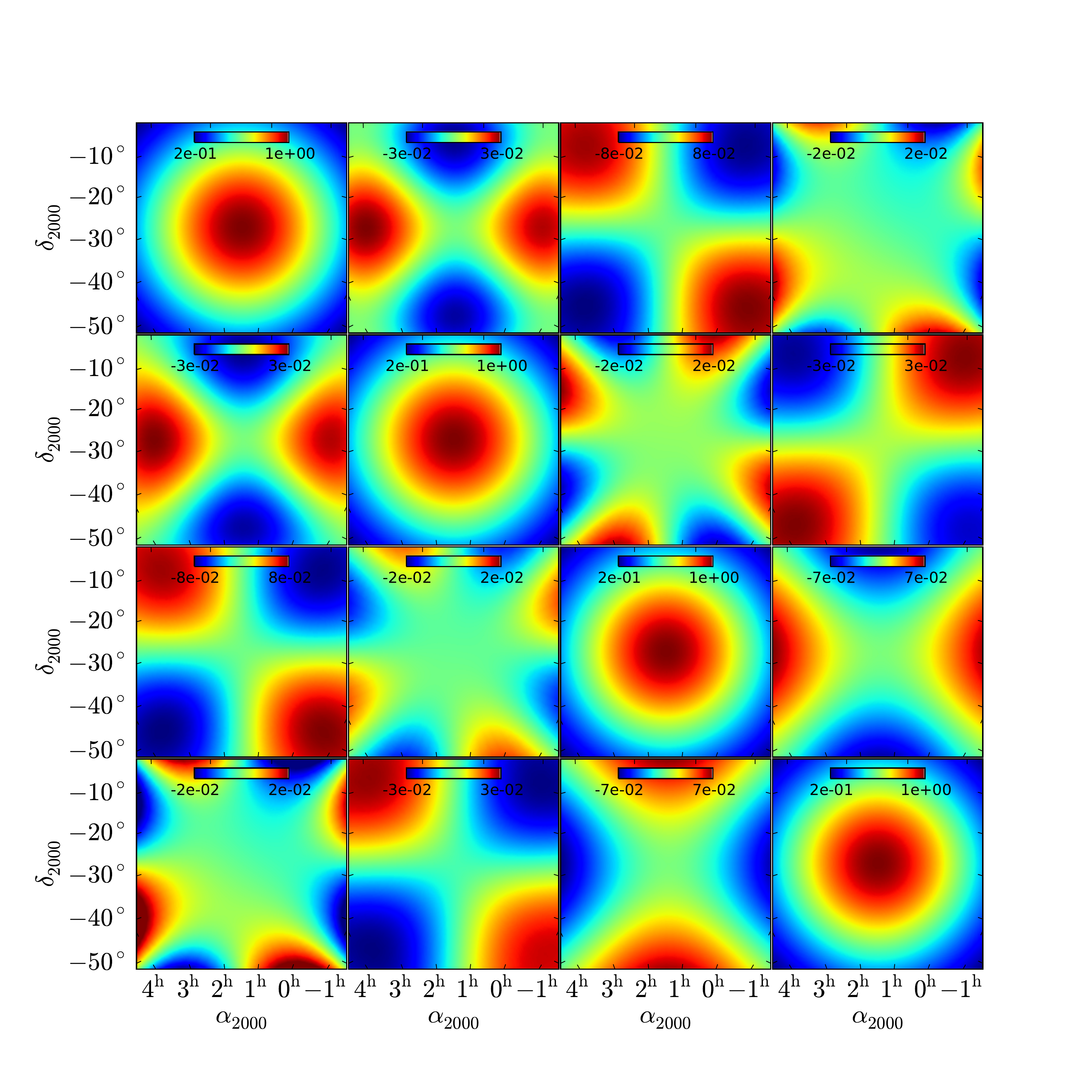}
\caption{${\bf A}$ matrices simulated at 130 (left) and 150~MHz (right) respectively. From the simulated complex dipole patterns, ${\bf A}$ matrices are computed in the altitude--azimuth coordinate system with the $x$ dipole assumed to be aligned East-West. Here they are resampled on an $(l,m)$ regular grid over a $45^{\circ}$ field of view, centered at an arbitrary right ascension $\alpha=1^{\rm h} 24^{\rm m}$ - the declination remains fixed at the telescope location $\delta = -30^\circ 43' 17''$. Resampling is carried out through a weighted average amongst the three closest points, with weights equals to the inverse distance to the $(l,m)$ grid point.}
\label{fig:Mueller_mat}
\end{figure*}

\begin{figure}[ht]
\centering
\includegraphics[width=1.1\columnwidth]{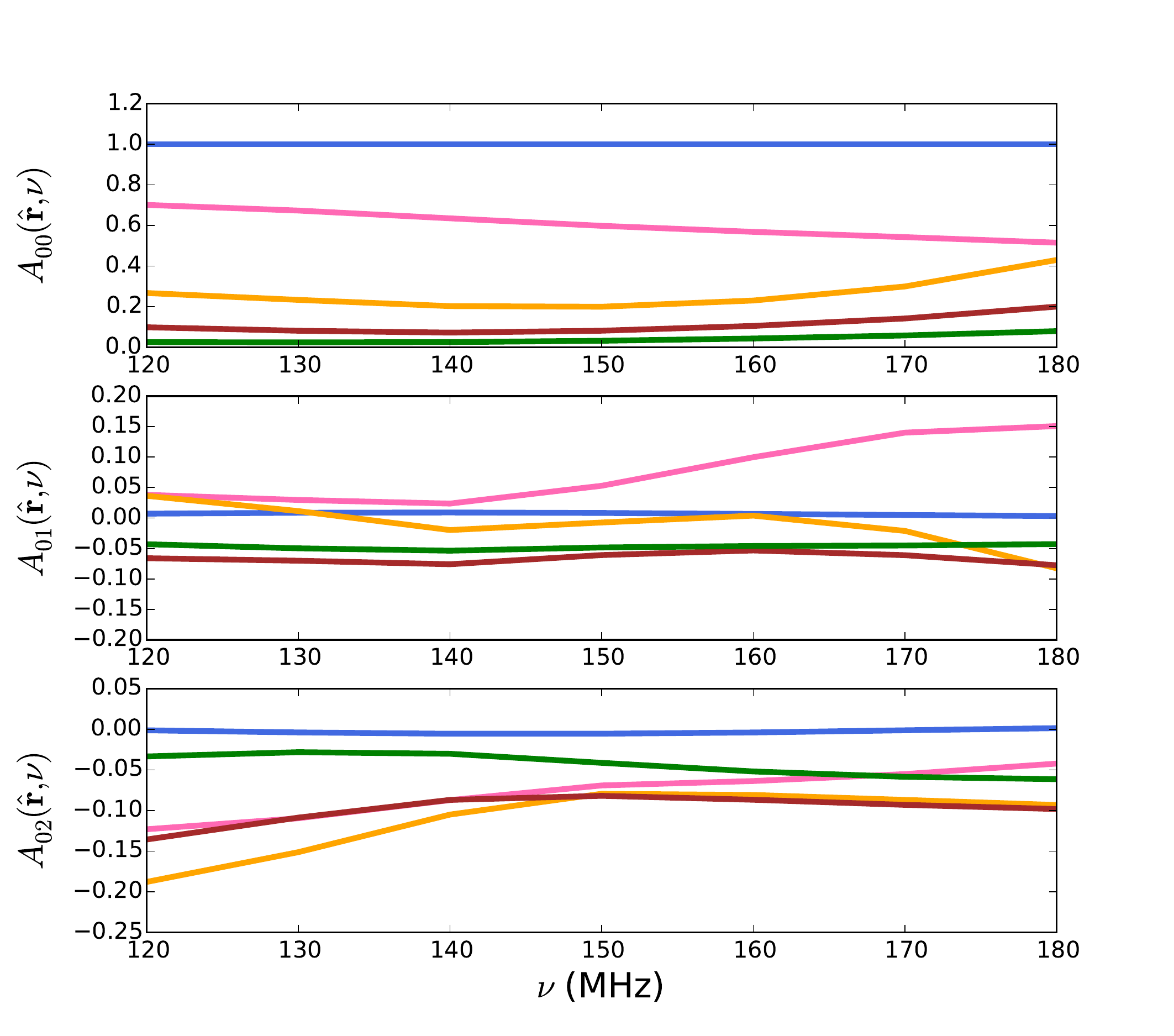}
\caption{Spectral behaviour of the ${\bf A}$ matrix (first row only) along a few selected lines of sight ${\bf {\hat r}}$ corresponding to a $20^\circ$ azimuth and $90^\circ$ (blue), $70^\circ$ (pink), $50^\circ$ (orange), $30^\circ$ (brown) and $10^\circ$ (green) altitude respectively. The $A_{00}(\hat {\bf r},\nu)$ term is normalized to unity at zenith by construction. The $A_{03}(\hat {\bf r},\nu)$ term is not included as no Stokes~V sky emission is assumed throughout the paper.}
\label{fig:beam_freq}
\end{figure}

We used the FEKO\footnote{\url{https://www.feko.info/product-detail}} package in order to obtain a model of the PAPER x--y complex dipole patterns in the $100-200$~MHz range, spaced 10~MHz apart. The model is based on the dipole physical dimensions and includes a reflective mesh positioned above the ground. Examples of the corresponding ${\bf A}(\hat{\boldsymbol{r}}, \nu)$ matrices are shown in Figure~\ref{fig:Mueller_mat} whereas Figure~\ref{fig:beam_freq} shows the frequency behaviour of the first row of ${\bf A}(\hat{\boldsymbol{r}}, \nu)$ (first three elements) for a few selected lines of sight. The $A_{00}(\hat {\bf r},\nu)$ term has a smooth frequency behaviour that decreases slowly and monotonically from zenith to the horizon. This behaviour is qualitatively in agreement with the beam model presented in \cite{Parsons2010}, although we defer a more quantitative comparison to future work. The off-diagonal terms representing the leakage to Stokes~I have a more complex spatial and spectral behaviour: their magnitude is essentially negligible at zenith whereas it becomes more than 10\% already at $70^\circ$ altitude.

\begin{figure}[ht]
\centering
\includegraphics[width=1.0\columnwidth]{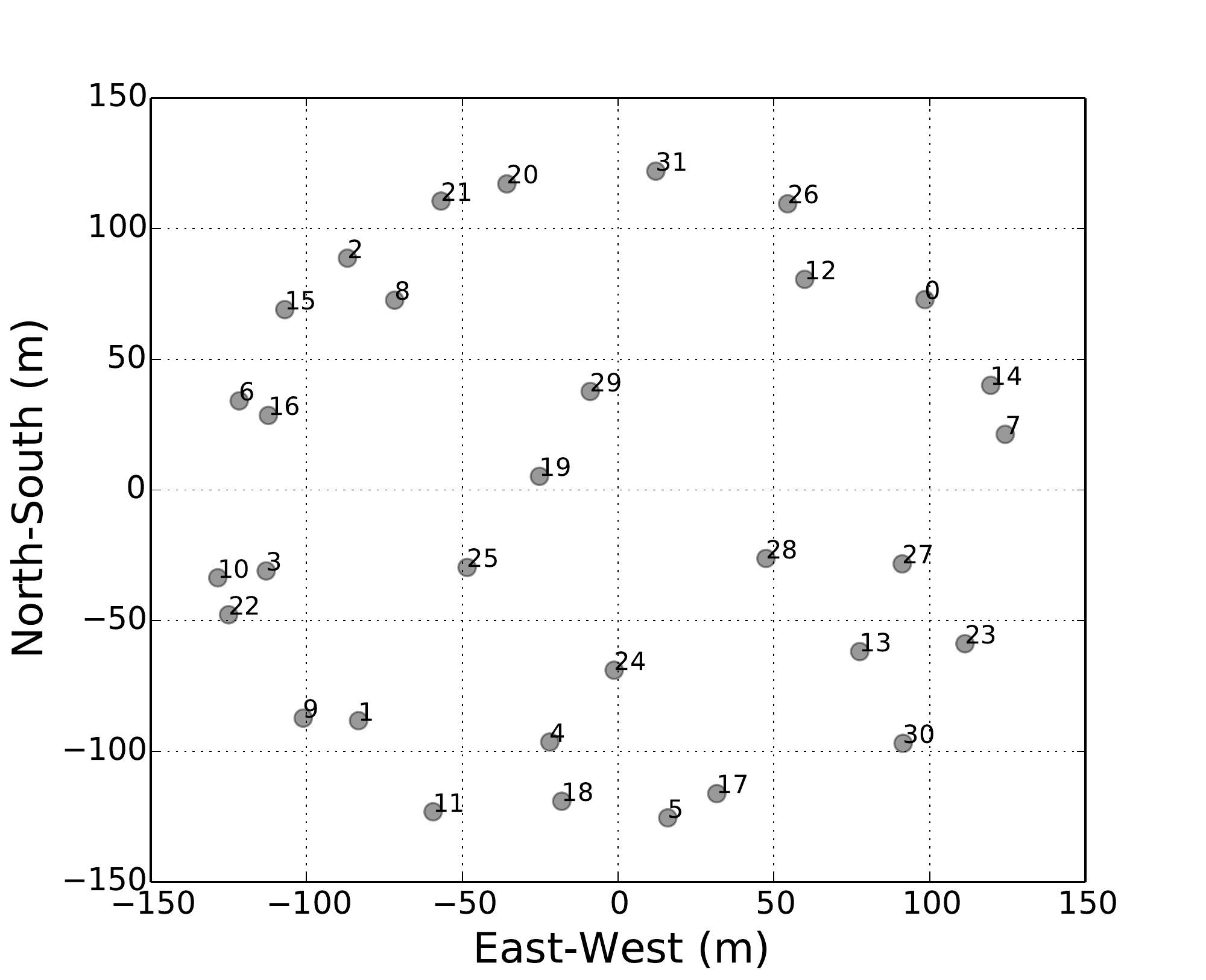}
\caption{Array layout used for our simulations.}
\label{fig:config}
\end{figure}

We used the configuration of the PAPER 32-element imaging array \citep[][Figure~\ref{fig:config}]{Stefan2013,Jacobs2013,Kohn2016} to simulate a baseline distribution that covers a relatively wide range of $k_\perp$ values while retaining the 30~m baseline sample used in the estimate of the 21-cm power spectrum \citep[][]{Parsons2014,Jacobs2015,Ali2015}. 

We simulated a drift scan observation corresponding to actual PAPER 21-cm observations, i.e. spanning the $0^{\rm h} < {\rm LST} < 8.5^{\rm h}$ range with a 10~minute cadence, covering the $120-180$~MHz frequency range with 500~kHz wide channels. For each frequency channel, the simulated ${\bf A}$ matrices were obtained by averaging over the two closest frequencies at which the FEKO simulations were carried out.

For each time sample, frequency channel and baseline we performed the following operations:
\begin{enumerate}
	\item  generated the sky emission evaluating equation~\ref{eq:intfresp}, where the sky model consists of either a catalogue of point sources or a realization of diffuse Galactic emission (see Sections~\ref{sec:point_sources} and~\ref{sec:diffuse_emission} below);
	\item  simulated visibilities via a discrete Fourier transform implementation of equation~\ref{eq:visibility_2};
	\item  generated delay--transformed visibilities via a fast Fourier transform implementation of equation~\ref{eq:delay_3} where we used a Blackmann--Harris window function;
	\item  computed power spectra using equation~\ref{eq:power_spectrum_2}.
\end{enumerate}
In the following sections we describe the input foreground models to our simulations.

\subsection{Point source all-sky model}
\label{sec:point_sources}
Simulations of polarized point sources are completely defined by a catalogue that includes both the polarization fraction and the rotation measure (RM) values. Stokes $Q$ and $U$ parameters at any frequency $\nu$, for any source $i$ can indeed be computed as:
\begin{eqnarray}\label{eq:stokesIQU}
Q _{\nu,i} = \gamma_i \, I_{\nu,i} \cos(2 \chi_i) = \gamma_i \, I_{\nu_0,i} \left ( \frac{\nu}{\nu_0} \right )^{\alpha_i} \cos{\left ( 2 \, {\rm RM}_i \frac{c^2}{\nu^2} \right )} \nonumber \\
U_{\nu,i} = \gamma_i \, I_{\nu,i} \sin(2 \chi_i) = \gamma_i \, I_{\nu_0,i} \left ( \frac{\nu}{\nu_0} \right )^{\alpha_i} \sin{\left ( 2 \, {\rm RM}_i \frac{c^2}{\nu^2} \right )} \nonumber \\
\end{eqnarray}
where $\gamma$ is the polarization fraction, $I_{\nu_0}$ is the flux density at the $\nu_0$ reference frequency and $\alpha$ is the spectral index. We note that the knowledge of the absolute polarization angle is not necessary for the purpose of estimating the 21-cm power spectrum leakage, therefore we set it to be zero along each line of sight.

For the total intensity properties we used the \cite{HurleyWalker2014} catalogue that lists all the sources brighter than 120~mJy at 150~MHz and covers the Southern Hemisphere at $-58^{\circ}< \delta < -14^\circ$. Although this catalogue is only somewhat deep, it has the advantage of providing the actual source locations, flux densities and spectral indexes in the $100-200$~MHz band.
From the \cite{HurleyWalker2014} catalogue it is possible to generate a polarized catalogue assuming the RM and polarization fraction statistics respectively. Detailed, wide--area information on the low--frequency polarization properties of radio sources are still lacking to date, therefore we constructed a polarized catalogue derived from statical properties measured at higher frequencies. 

\begin{figure}[ht]
\centering
\includegraphics[width=1.0\columnwidth]{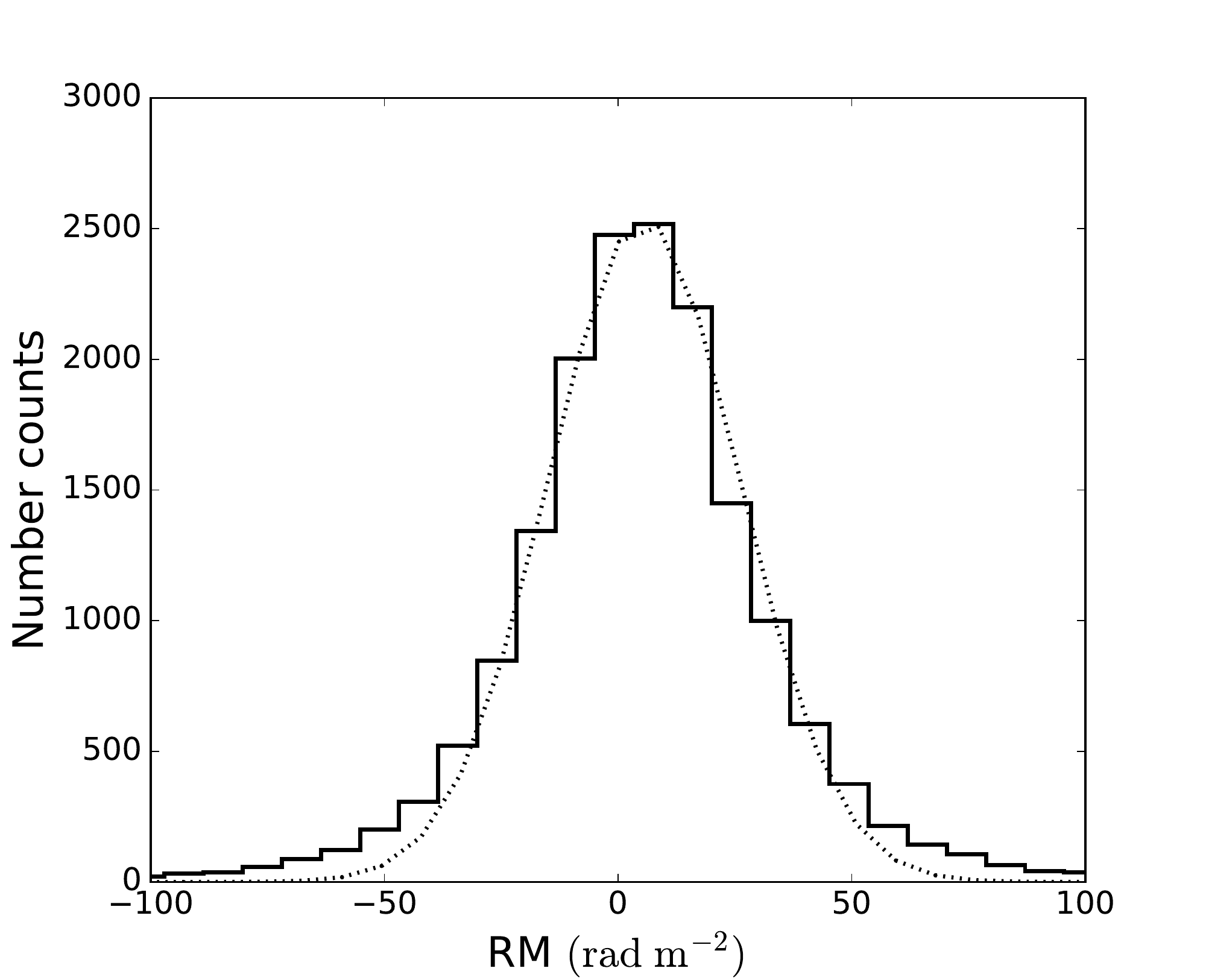}
%\vspace{0.005cm}
\caption{The RM distribution for sources at Galactic latitude $|b| > 20^{\circ}$ from the \cite{Taylor2009} catalogue. The dotted line shows the Gaussian best fit with a 5.6~rad~m$^{-2}$ mean and a 20~rad~m$^{-2}$ standard deviation.}
\vspace{0.1cm}
\label{fig:RM_dist}
\end{figure}

The RM distribution was taken from the 1.4~GHz catalogue by \cite{Taylor2009}, one of the most comprehensive polarization catalogues to date. For high Galactic latitude sources ($|b| > 20^\circ$), the RM distribution fairly follows a Gaussian profile, with values as high as $\sim |100|$~rad~m$^{-2}$ (Figure~\ref{fig:RM_dist}). As the rotation measure is the integral of the magnetic field along the line of sight weighted by the electron density, we do not expect it to change with frequency, therefore, for each simulated source, we assigned an RM value drawn from the Gaussian best fit to the RM distribution of the high Galactic latitude sources in the \cite{Taylor2009} catalogue.

The statistics of the polarization fraction $\gamma$ is more uncertain as is, instead, expected to decrease with frequency due to internal Faraday dispersion \citep{Burn1966}. It is not, therefore, straightforward to extrapolate the average polarization fraction at 1.4~GHz as Faraday depolarization depends on the specific source physical conditions \citep[e.g., geometry and magnetic field strength,][]{Tribble1991}. 
Recent observations \citep[][]{Bernardi2013,Mulcahy2014,Asad2016} have indeed started to show that the average polarization fraction of radio sources decreases from a few percent value at 1.4~GHz to less than 1\% at 150~MHz. \cite{Lenc2016} present the most stringent constraints to date to be $\langle \gamma \rangle \le 0.32\%$ at 154~MHz, derived from a sample of 187 sources. We conservatively treated this result as a measurement rather than an upper limit and assigned a polarization fraction value drawn from a uniform distribution between 0 and 0.32\% to each simulated source.

\subsection{Galactic diffuse emission model}
\label{sec:diffuse_emission}
Observations of diffuse, Galactic polarized emission reveal a wealth of spatial structures in the interstellar medium (ISM) that are strongly frequency dependent. In particular, recalling that the Faraday depth $\phi$ of a synchrotron emitting region between two points $l_1$ and $l_2$ along an arbitrary line of sight $\hat{\boldsymbol{r}}$, given the electron density $n_e$ and magnetic field along the line of sight $B_\parallel$ is defined as \citep[][]{Burn1966,Brentjens2005}:
\begin{eqnarray}\label{eq:faraday_depth}
\phi({\hat{\boldsymbol{r}}},l_1,l_2) = 0.81 \int_{l_1}^{l_2} n_e B_\parallel dl,
\end{eqnarray}
observations below a few hundred MHz show multiple polarized structures in the ISM at different Faraday depth values for almost any given line of sight. \cite{Jelic2010} and \cite{Alonso2014} have recently attempted to model this complexity, although a realistic description requires the knowledge of the distribution of the polarized emission both on angular scales and in Faraday depth, which is only partially emerging from recent observations \citep{Bernardi2013,Jelic2014,Jelic2015,Lenc2016}.

In this work, we made the simplifying assumption to ignore the spatial and line--of--sight Faraday depth structure and consider the contribution from two representative Faraday depths integrated all the way to the observer's location. With this approximation, the Faraday depth coincides with the RM and equation~\ref{eq:stokesIQU} can be used to obtain Stokes $Q$ and $U$ all--sky maps too:
\begin{eqnarray}\label{eq:stokesQU-diffuse}
Q ({\hat{\boldsymbol{r}}, \nu}) & = & P(\hat{\boldsymbol{r}}) \cos{\left ( 2 \, \phi  \frac{c^2}{\nu^2} \right )} \nonumber \\
U ({\hat{\boldsymbol{r}}, \nu}) & = & P(\hat{\boldsymbol{r}}) \sin{\left ( 2 \, \phi  \frac{c^2}{\nu^2} \right )}.
\end{eqnarray}
The all--sky, polarized intensity map $P$ could be, in principle, derived from a total intensity one, however, interferometric observations of polarized Galactic emission at all radio frequencies are known to suffer from the so called ``missing short spacing problem": they filter out the large scale, smooth background emission and retain the small scale, Faraday--rotated foreground structures introduced by local ISM fluctuations in either the electron density or the magnetic field \citep[e.g.,][]{wieringa1993,Gaensler2011}. This effect leads to the lack of correlation between total intensity and polarized diffuse emission, with an apparent polarization percentage exceeding 100\% \citep[e.g.][]{Gaensler2001,haverkorn2003,Bernardi2003,schnitzeler2009,Iacobelli2013,Bernardi2010}, preventing from using a total intensity template for polarization simulations. 

In order to overcome this problem, we follow an approach similar to \cite{Alonso2014}, who simulated a polarized foreground map $P(\hat{\boldsymbol{r}})$ at a reference frequency $\nu_0$ from the polarized spatial power spectrum $C_{\ell}^P$:
\begin{equation}\label{eq:PI-pspectrum-def}
\langle \tilde{P} ({\bf l}) \tilde{P} ({\bf l})^{*} \rangle = (2 \pi)^2 C_{\ell}^P \delta^{(2)} ({\bf l} - {\bf l}')
\end{equation}
where $\tilde{P}$ is the Fourier transform of $P$, ${\bf l}$ is the two dimensional coordinate in Fourier space, $\langle \, \rangle$ is the ensemble average, $\ell = \frac{180}{\Theta}$ with $\Theta$ the angular scale in degrees and $\delta^{(2)}$ is the two--dimensional Dirac function. 

The synchrotron polarized power spectrum obtained from large--area, GHz--frequency surveys is well described by a power law \citep[e.g.,][]{LaPorta2006,Carretti2010}:
\begin{equation}\label{eq:PI-pspectrum}
C_{\ell}^P = A_{\ell_0}^P \Big( \frac{\ell}{\ell_0} \Big)^{-\beta^P},
\end{equation}
down to $\ell \sim 100-1000$, with $2 < \beta^P < 3$. Due to the strong frequency dependence of the polarized emission \citep[e.g.][]{Carretti2005}, the extrapolation to low frequencies is very uncertain. 

\cite{Bernardi2009} and \cite{Jelic2015} measured the polarized spatial power spectrum at 150~MHz and found it to follow a power law with $\beta^P = -1.65$ in the $100 \le \ell \le 2700$ range, somewhat flatter than the higher frequency values. We therefore adopted this value for our simulations. As they both observed a relatively small ($6^\circ \times 6^\circ$) sky patch, their measurement of the power spectrum amplitude $A_{\ell_0}^P$ may be sample variance limited, therefore we
constrained it by using the 2400~square~degree survey carried out at 189~MHz by \cite{Bernardi2013}. 
The survey shows significant difference in the levels of polarized emission as a function of Galactic latitude, with maximum emission around the south Galactic pole and mostly concentrated at $|\phi| < 12$~rad~m$^{-2}$. 
We identified the $20^\circ \times 20^\circ$ area centered at $(l,b) \sim (200^{\circ},-80^{\circ})$ to be the brightest polarized emission region in the survey and selected the $\phi = 6, 12$~rad~m$^{-2}$ frames as the brightest frame and the representative of a typical high $\phi$ value for diffuse emission, in good agreement with the distribution of Faraday depth peaks recently measured in the Southern Galactic pole area at 154~MHz by \cite{Lenc2016}. 
We labeled the models corresponding to these two frames as $D6$ and $D12$ and measured their polarized intensity root--mean--square (rms) to be $P_{{\rm rms},D6} =1$~K and $P_{{\rm rms},D12} = 0.21$~K at $\phi_{D6} = 6$~rad~m$^{-2}$ and $\phi_{D12} = 12$~rad~m$^{-2}$ respectively. 
The power spectrum amplitude $A_{D6,D12}$ was obtained through its relationship with the measured rms value \citep[e.g.,][]{zaldarriaga2004}:
\begin{eqnarray}\label{eq:Trms}
P_{{\rm rms},D6,D12}	& = & \sqrt{\sum_{\ell_1}^{\ell_2} \frac{(2 \ell + 1)}{4\pi}C_{\ell,D6,D12}^P}  \nonumber \\
			& = & \sqrt{\sum_{\ell_1}^{\ell_2} \frac{(2 \ell + 1)}{4\pi} A_{D6,D12}  \Big(\frac{\ell}{\ell_0} \Big)^{-\beta^P}},
\end{eqnarray}
where we dropped the $\ell_0$ subscript for clarity and the exclusion of short baselines and the angular resolution of the \cite{Bernardi2013} survey set $\ell_1 \sim 100$ and $\ell_2 \sim 680$ respectively. 

At this point we have all the necessary ingredients to generate the two $D6$ and $D12$ realizations of the diffuse polarized emission model using equation~\ref{eq:stokesQU-diffuse}. We generated a map $P_{D6} (\hat{\boldsymbol{r}})$ ($P_{D12} (\hat{\boldsymbol{r}})$) from the polarized power spectrum $C_{\ell,{D6}}^P$ ($C_{\ell,{D12}}^P$) using the Healpix \citep{Gorski2005} routine SYNFAST. We chose an ${\rm N}_{\rm side} = 128$ parameter that retains sufficient sampling for the 30~m baseline used for power spectrum estimation. We then substituted $\phi_{D6}$ ($\phi_{D12}$) in equation~\ref{eq:stokesQU-diffuse} to generate the Stokes $Q$ and $U$ full--sky maps corresponding to the $D6$ ($D12$) model realization.

\section{Results}
\label{sec:results}
In this section, we compare our simulations with existing observations \citep{Kohn2016}, following which 
we predict the leakage expected in 21-cm power spectrum measurements. We then provide constraints on the average point source polarization fraction based on the data from \citet{Moore2015}.

\subsection{Polarized Power Spectra}
\label{subsec:unpolarized power_spectra}
\begin{figure*}
\centering
\includegraphics[width=2\columnwidth]{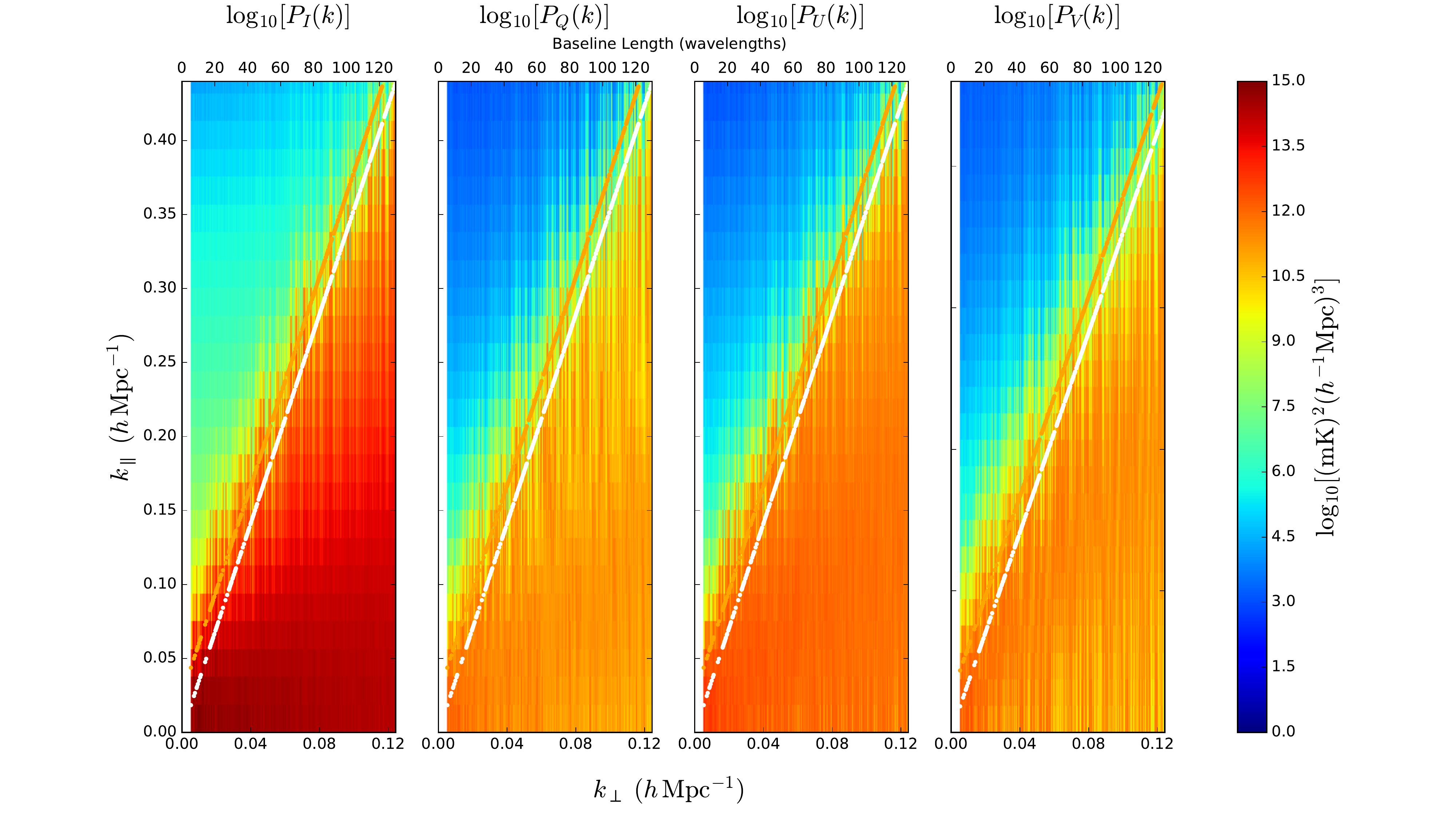}
\caption{Two--dimensional power spectra ${\bf p} = (p_I,p_Q,p_U,p_V)$ (left to right) over a 25~MHz bandwidth centered at 150~MHz, obtained from the brightest 1500 unpolarized total intensity sources from the \cite{HurleyWalker2014} catalogue. The simulation only includes Stokes I; Stokes Q, U and V are leakages due to the instrumental widefield effects. The white line marks the horizon limit and the orange line is 50~ns beyond.}
\label{fig:MWAwedges_n1500-unpol}
\end{figure*}
\begin{figure}
\includegraphics[width=1\columnwidth]{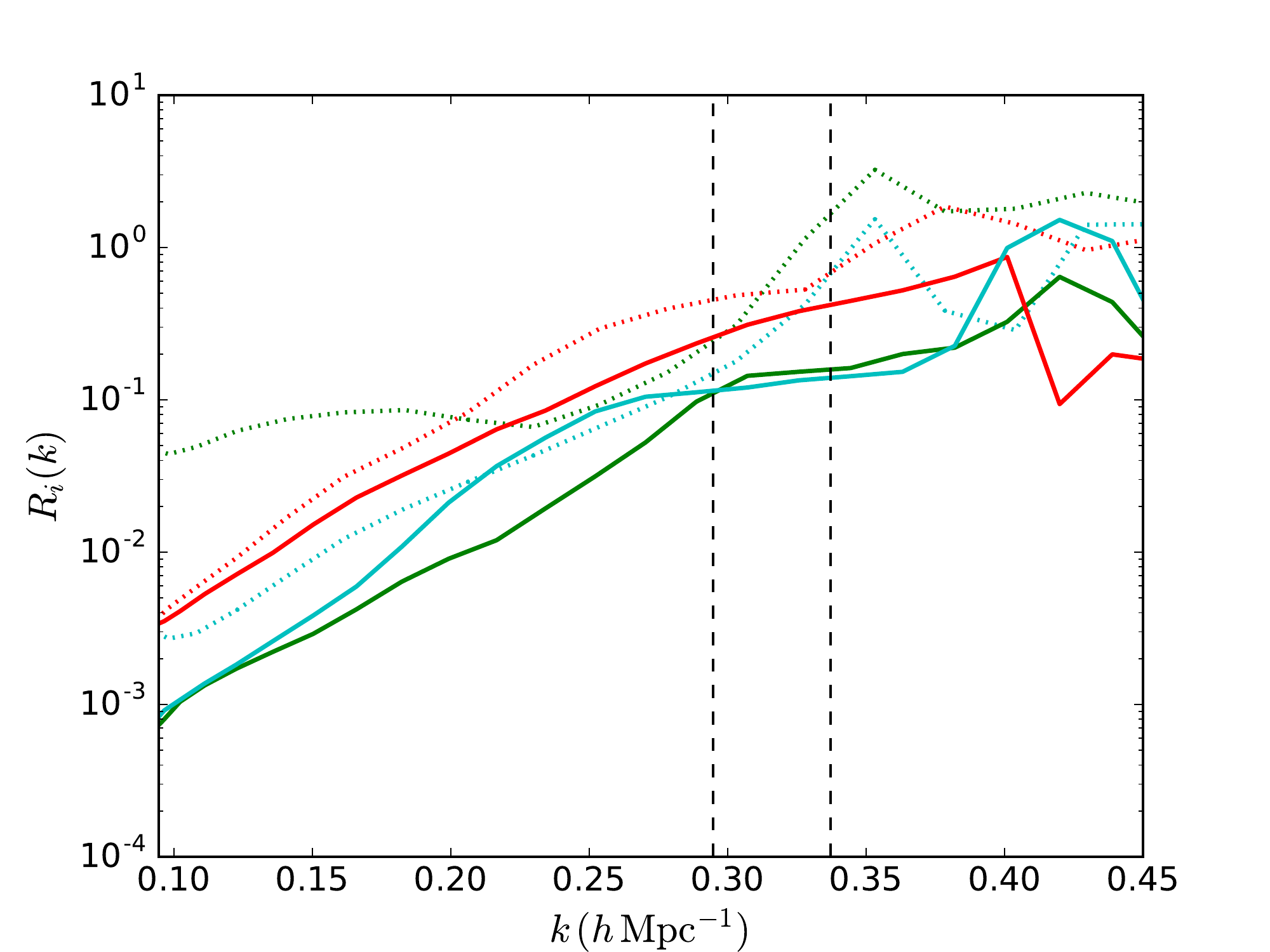}
\caption{Ratios of Stokes~$Q$ (solid green), $U$ (solid red) and $V$ (solid cyan) versus $I$ power spectra averaged over a $0.092<k_{\perp}<0.097\,h$~Mpc$^{-1}$ region, whose centre corresponds to a 175~m baseline. The dotted lines represent the corresponding power spectrum ratios from Figure~7 in \cite{Kohn2016}. The left and right vertical dashed lines mark the horizon limit and 50~ns beyond it respectively.} 
\label{fig:SphericalPSvsSaul-unpol}
\end{figure}
We first tested our simulation framework with a sky model composed of total intensity sources from the \cite{HurleyWalker2014} catalogue. In this case we expected to reproduce the well established two--dimensional wedge--like total intensity power spectrum $p_I$ observed, for example, in \cite{Pober2013}, \cite{Thyagarajan2015b} and \cite{Kohn2016}. Figure~\ref{fig:MWAwedges_n1500-unpol} displays power spectra from this simulation using only the 1500 brightest sources down to a flux density threshold of 2~Jy. Power spectra change by only a few percent with the inclusion of all the sources down to the 120~mJy catalogue threshold, therefore the decreased computing load in using only the 1500 brightest sources merits the small loss in accuracy for the aim of this paper.

The simulated total intensity power spectrum $p_I$ shows the wedge--like morphology and power levels similar to \cite{Pober2013}, confirming that our formalism is consistent with previous works that are limited to the total intensity case. Polarized power spectra display, in this case, the leakage from total intensity due to widefield polarized primary beams. All polarized power spectra have a similar behavior, with emission confined in a wedge--like shape very similar to total intensity. The power ratio between the emission inside and outside the wedge is at the $10^9-10^{10}$ level, indicating that very little
chromatic structure has been introduced by the primary beam outside the wedge. 
	
In order to provide a first order validation of the beam models used in simulations, we compared the ratio $R_i$ of polarized versus total intensity power spectra defined as
\begin{equation}\label{eq:beamerror}
R_i (k) = \frac{p_i (k)}{p_I (k)}, \, \, \, \rm{with } \, \, i = Q, U, V
\end{equation}
calculated from our simulations and from power spectra measured in a 5~hour transit observation with PAPER \citep{Kohn2016}. The ratio is insensitive to a possible different absolute normalization between the simulation and data. Figure~\ref{fig:SphericalPSvsSaul-unpol} shows that simulated $R_i$ ratios are generally fainter than the measured ones, although the simulated $R_U$ and $R_V$ substantially agree with the measured ones. The measured power spectra show emission that extends up to 50~ns beyond the horizon limit, after which they appear to be noise-dominated \citep{Kohn2016}. As our simulations essentially have no emission beyond the horizon limit, this explains why both simulated and measured $R_i$ approach unity outside the wedge.
The largest difference between simulations and data appears in $R_Q$ at $k < 0.2 \, h$~Mpc$^{-1}$, where our model underpredicts the measured value by more than one order of magnitude. This is the consequence of an excess of power found in $p_Q$ at small $k$ values by \citet{Kohn2016} that may be attributable to a calibration mismatch of the two orthogonal polarizations and, possibly, to intrinsic sky polarization not included in our simulations.

\subsection{Predictions of polarization leakage}
\label{subsec:polarized_spectra}
\begin{figure}
\centering
\includegraphics[width=1\columnwidth]{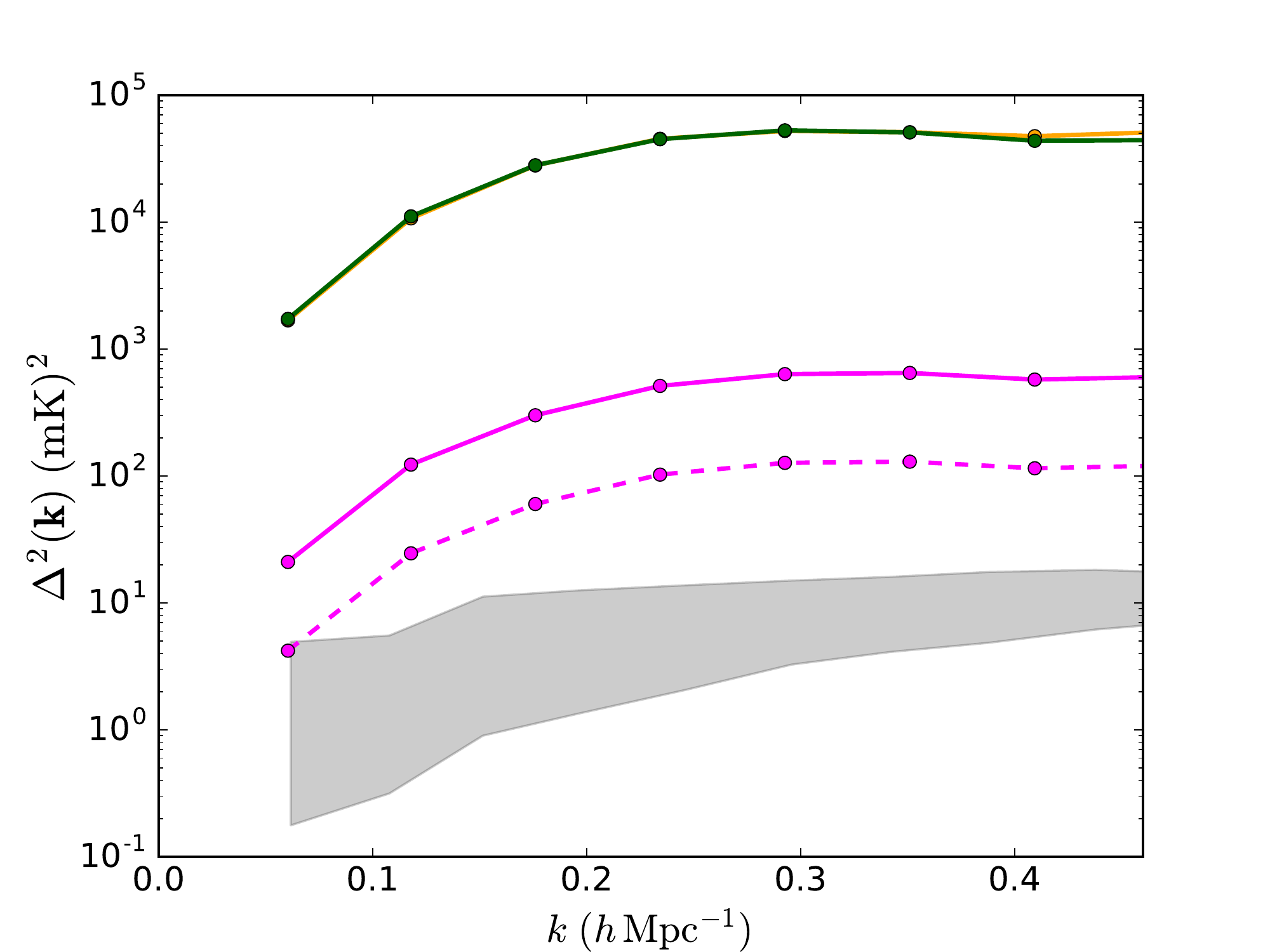}
\caption{Polarized power spectra $\Delta^2_{Q,U}$ (orange and green lines respectively) for a 30~m baseline and an 8~MHz bandwidth centered at 150~MHz ($z = 8.5$) from polarized point source sky model with the total intensity intentionally set to zero (see text for details). The magenta line represents $\Delta_I^2$, i.e. the predicted leakage to total intensity. The magenta dashed line shows the leakage when we assumed the polarization fraction to be distributed between 0 and 0.14\%. The shaded gray region represents power spectra of the 21-cm fiducial model from \cite{Lidz2008} with HI neutral fractions ranging between 0.21 and 0.82.}
\label{fig:SphericallPS_150MHz}
\end{figure}

\begin{figure*}
\centering
\includegraphics[width=1\columnwidth]{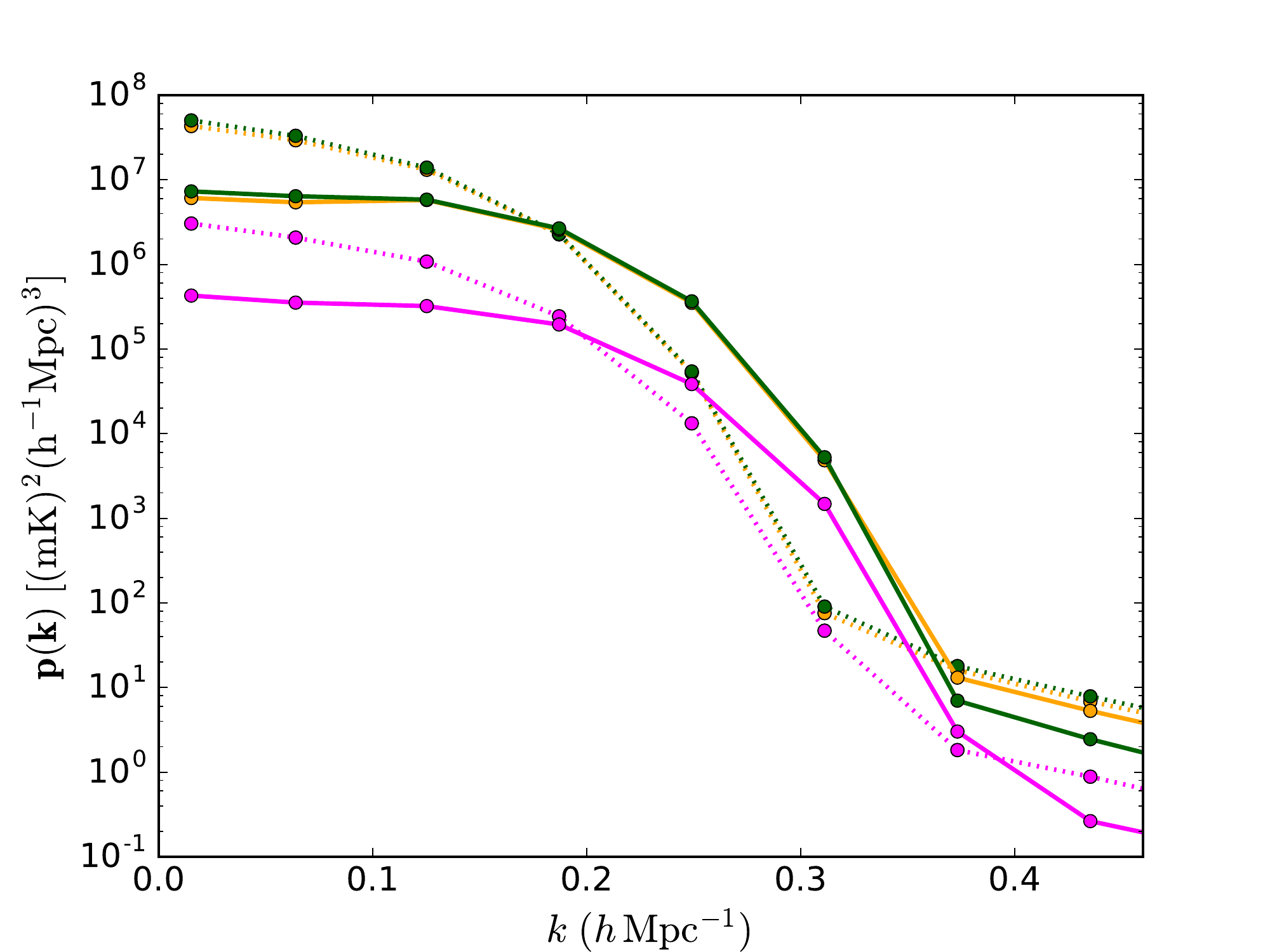}
\includegraphics[width=1\columnwidth]{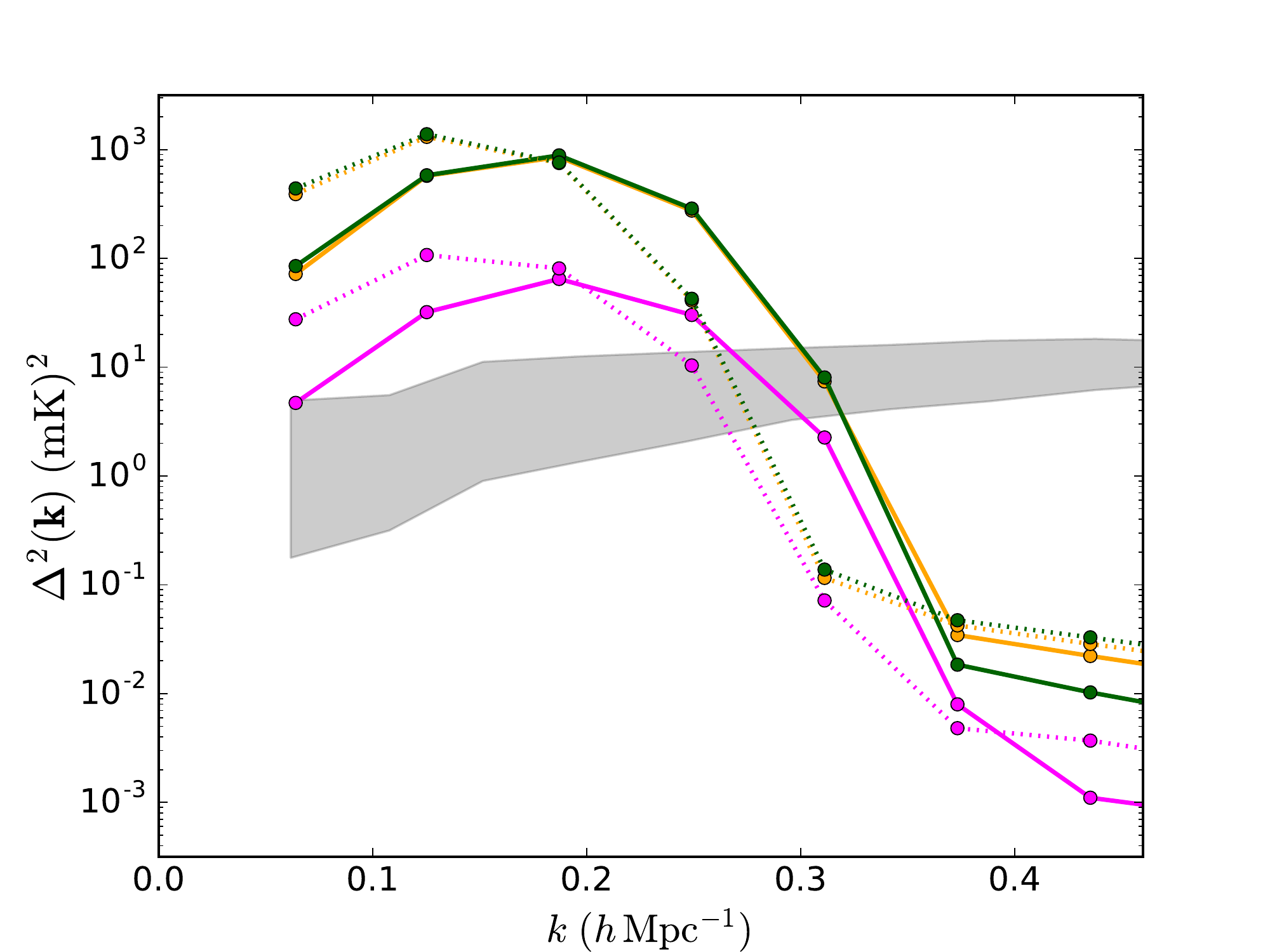}
\caption{\textit{Left panel:} Polarized power spectra $P_{Q,U}$ (orange and green lines respectively) for a 30~m baseline and an 8~MHz bandwidth centered at 150~MHz ($z = 8.5$) from the diffuse polarized foreground $D12$ (solid) and $D6$ (dotted). The magenta line represents the leaked power spectra $P_I$. \textit{Right panel:} Same as left panel but for $\Delta^2_{I,Q,U}(k)$. The shaded gray region represents power spectra of the 21-cm fiducial model from \cite{Lidz2008} with HI neutral fractions ranging between 0.21 and 0.82.}
\label{fig:SphericalPS-Diffuse}
\end{figure*}

The simulation framework developed in this paper eventually aims to predict the amount of leakage expected in the measured 21-cm power spectra. In order to do so, we carried out two sets of simulations for the models described in Sections~\ref{sec:point_sources} and \ref{sec:diffuse_emission} respectively, where we explicitly set the point source total intensity model to zero, so that $p_I$ directly measures the leakage from polarized foregrounds. The resulting power spectra at $z = 8.5$ (150~MHz) for a fiducial 30~m baseline are shown in Figure~\ref{fig:SphericallPS_150MHz} and \ref{fig:SphericalPS-Diffuse} respectively.

Polarized power spectra generated from the point source simulation exhibits significant super--horizon emission, i.e. almost constant power outside the $k \sim 0.06\,h$~Mpc$^{-1}$ horizon limit for a 30~m baseline (Figure~\ref{fig:SphericallPS_150MHz}). This behavior is expected from Faraday rotated foregrounds whose leakage is not confined in $k_\parallel$ space. Power spectra also appear featureless in $k$ space and this can be intuitively understood as a single RM value corresponds to a specific $k_\parallel$ value \citep{Moore2015}:
\begin{eqnarray}\label{eq:kpRM}
k_{\parallel} = \frac{4\, \lambda^2\,H(z)}{c \, (1+z)}\,\textrm{RM}, 
\end{eqnarray}
therefore, a population of point sources distributed over a broad range of RM values essentially displays power at any $k_\parallel$ value. This is the reason why diffuse emission power spectra show a characteristic knee--shape as a function of $k$ (Figure~\ref{fig:SphericalPS-Diffuse}): by construction, they only have structure at $\phi = 6$ and 12~rad~m$^{-2}$ corresponding to $k_\parallel \approx 0.02$ and $0.06 \, h$~Mpc$^{-1}$ respectively and their power, therefore, falls off at higher $k_\parallel$ values. Noticeably, the $D12$ model power spectrum is brighter than the $D6$ one at $k > 0.2\, h$~Mpc$^{-1}$ despite its normalization being five times smaller (see Section~\ref{sec:diffuse_emission}), showing that the super--horizon contamination depends more on the $\phi$ value rather than the intrinsic foreground brightness. 

In terms of contamination to the 21-cm power spectrum, our predictions should be regarded as worst case scenarios, due to our conservative model assumptions. In both foreground models, leaked power spectra approximately behave as scaled versions of polarized power spectra, however, in the diffuse emission case, the leaked power spectrum is $\sim 0.03$~(mK)$^2$ for $0.3 < k < 0.5 \, h$~Mpc$^{-1}$, a reasonably negligible contamination to the 21-cm power spectrum. Bright, diffuse polarized foregrounds are therefore not a concerning contamination to the 21-cm power spectrum as long as their emission is confined at low Faraday depths as all the existing observations are showing \citep[e.g.][]{Bernardi2009,Bernardi2010,Bernardi2013,Jelic2014,Jelic2015,Lenc2016}.
From a pure avoidance perspective, therefore, knowledge of the polarized point source distribution is more relevant as it may contaminate high $k$ modes too: its leakage magnitude is a strong function of the average point source polarization fraction, becoming one order of magnitude smaller if a uniform distribution with a maximum value of 0.14\% is assumed (Figure~\ref{fig:SphericallPS_150MHz}, magenta dashed line). We will return on this point in the next section.

\begin{figure}
\centering
\includegraphics[width=1.05\columnwidth]{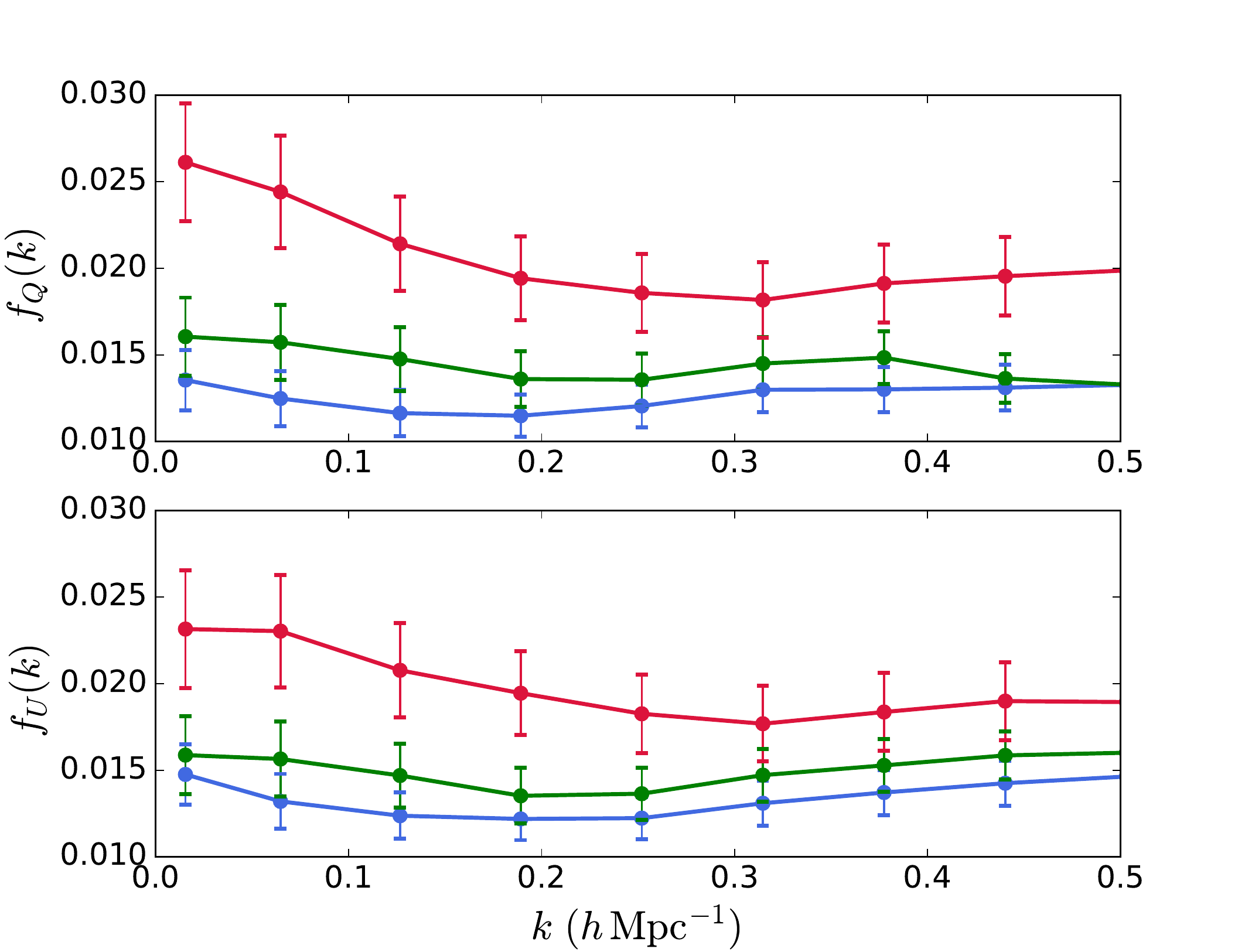}
\caption{Stokes~$Q$ (top panel) and $U$ (bottom panel) fractional leakage per $k$ mode for a 30~m baseline estimated from the polarized point source model at $z=8$ (blue line), $z=9$ (green line) and $z=10$ (red line) respectively. Error bars were calculated as the standard deviation of ten random realizations of the model.}
\label{fig:RatioI2QU}
\end{figure}

One natural by product of our formalism is the predicted fractional leakage $f$ per $k$ mode defined as the reciprocal of $R_i$:
\begin{equation}\label{eq:fractional-leakage}
f_i(k)=\frac{1}{R_i} = \frac{p_I (k)}{p_i (k)}
\end{equation}
Figure~\ref{fig:RatioI2QU} shows that the fractional leakage contributed by Stokes $Q$ and U is less that 3\% for $k < 0.5 \, h$~Mpc$^{-1}$ in the $8 < z < 10$ range and tends to increase with redshift. \cite{Moore2015} gave a simplified estimate of the fractional leakages that is consistent with ours within a factor of two.\\
Finally, we note that $f_i$ has a rather different behaviour as a function of $k$ than $R_i$, while they should be, in first approximation, similar, due to the fact that the ${\bf A}$ matrices are nearly symmetrical (Figure~\ref{fig:Mueller_mat}). There is an intrinsic difference due to the fact that the simulations presented here pertain to two different baselines, however, most of their difference is due to the input model. The unpolarized point source model (Figure~\ref{fig:SphericalPSvsSaul-unpol}) is very smooth in frequency by construction, leading to a very bright Stokes~$I$ power spectrum at small $k_\parallel$ values and, therefore, a corresponding $R_i$ at those modes; conversely, as mentioned above, the polarized point source model has power at essentially any $k_\parallel$ value by construction, leading to an almost flat $f_i$ as a function of $k_\parallel$.

\subsection{Constraining the polarization fraction}
\label{subsec:polarization_fraction}
\begin{figure*}
\centering
\includegraphics[width=1\columnwidth]{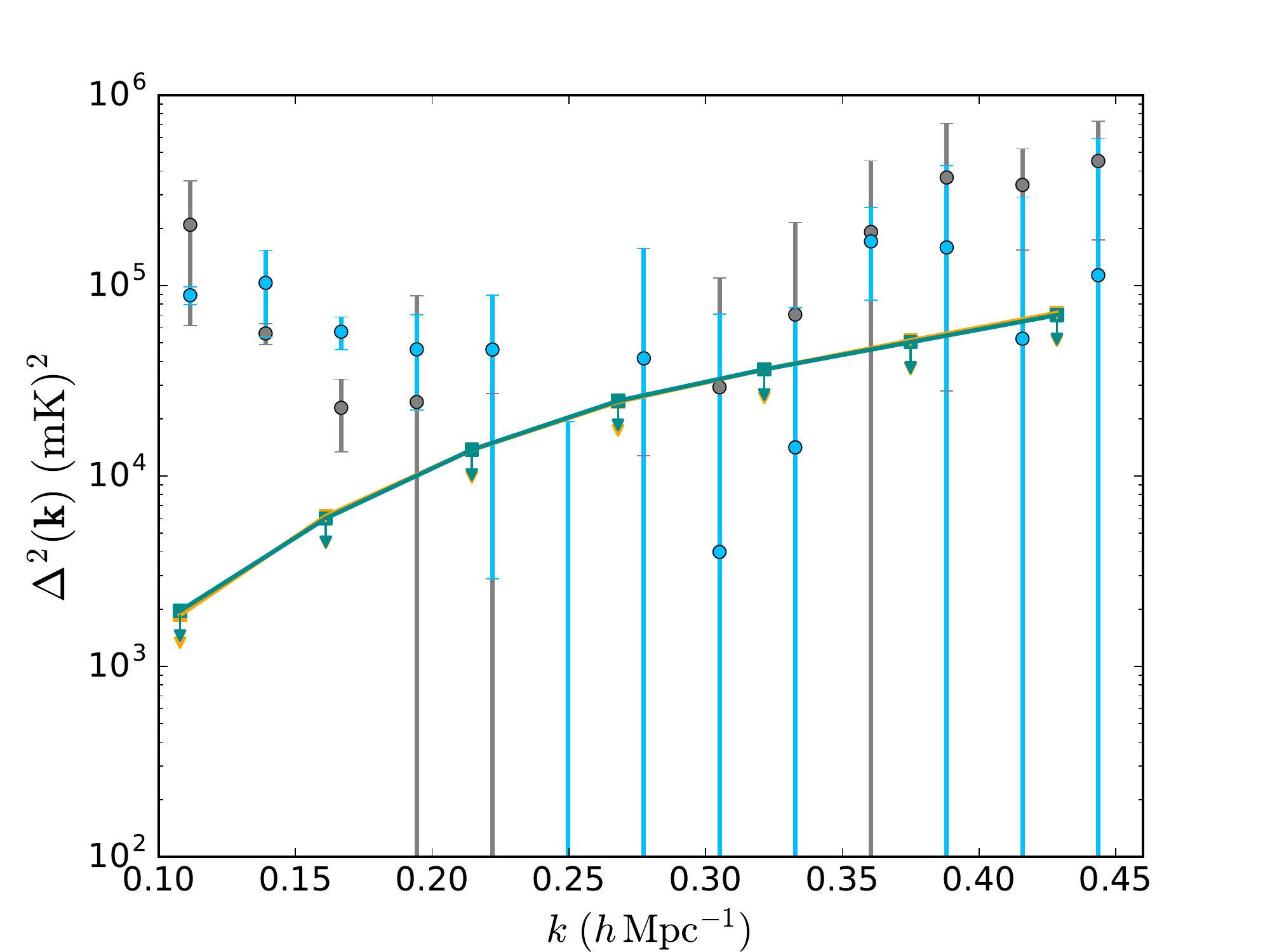}
\includegraphics[width=1\columnwidth]{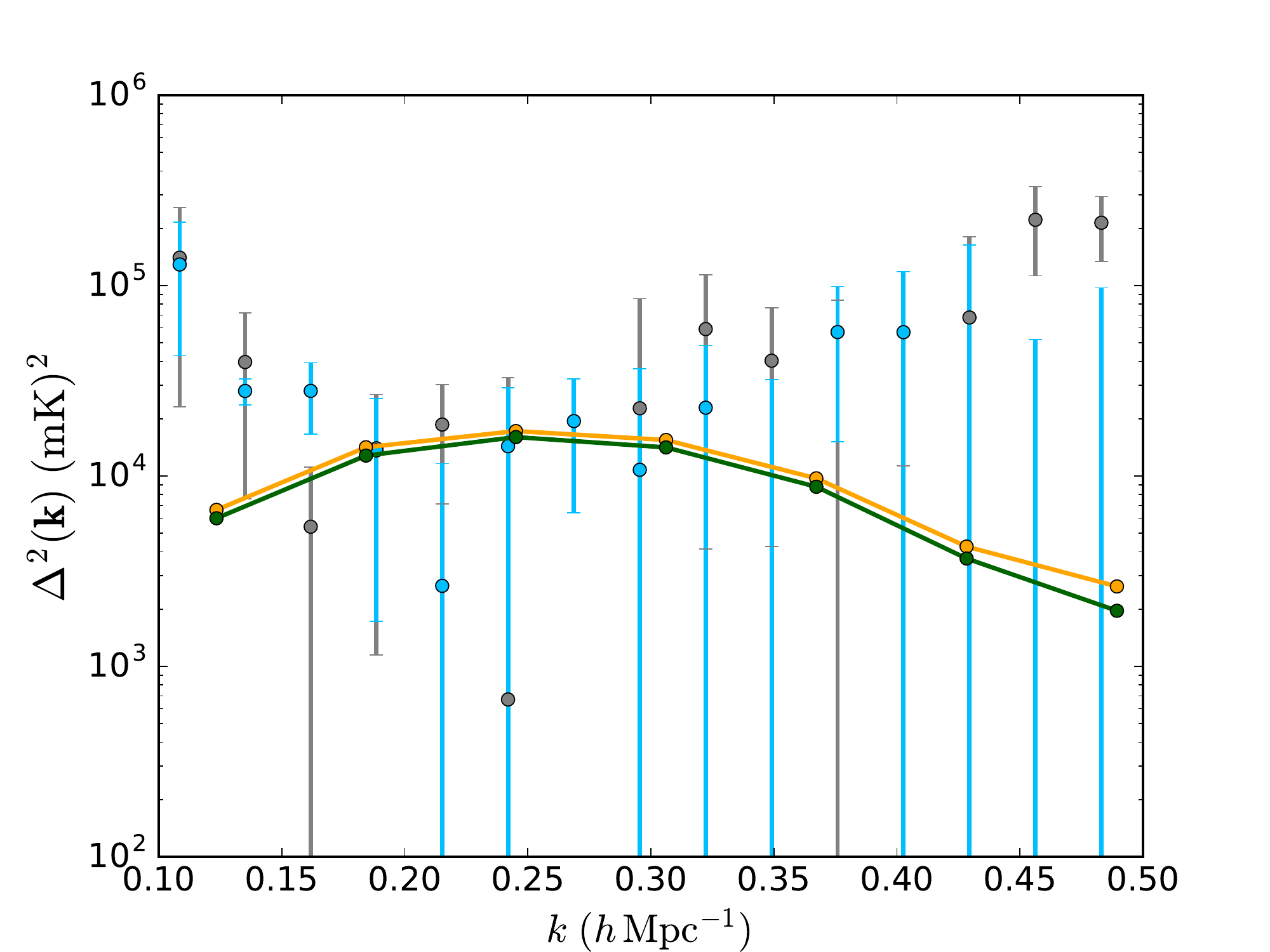}
\caption{\textit{Left panel:} Polarized power spectra $\Delta^2_{Q'}$ and $\Delta_{U'}^2$ (orange and green respectively) from our point source simulation scaled down to match the observed $\Delta^2_{Q_m}$ and $\Delta_{U_m}^2$ (blue and gray circles respectively) from \citet{Moore2015} at 126~MHz ($z=10.3$). \textit{Right panel:} Polarized power spectra $\Delta^2_{Q'}$ and $\Delta_{U'}^2$  predicted from our point source simulation at 164~MHz ($z = 7.7$) compared with  the observed $\Delta^2_{Q_m}$ and $\Delta_{U_m}^2$ (blue and gray circles respectively) from \citet{Moore2015} at the same frequency. We note that our predictions are compatible with the observed upper limits.}
\label{fig:SphericallPS-David}
\end{figure*}

We compared our predictions for the point source model - the worst expected contamination - with the polarized power spectrum measurements from a 30~m baseline deep integration with the PAPER--32 array \citep{Moore2015}. Their $\Delta^2_{Q_m}$ and $\Delta_{U_m}^2$ at 126 and 164~MHz (reported here in Figure~\ref{fig:SphericallPS-David}) are essentially consistent with the noise level in the $\Delta k = [0.2,0.45] \, h$~Mpc$^{-1}$ range. We used these results to constrain our polarized power spectrum $\Delta_{Q}^2$ to $\Delta_{Q'}^2$ at the same frequencies to be:
\begin{equation}
\Delta_{Q'}^2 = \frac{\langle \Delta^2_{Q,U} \rangle_{\Delta k}}{\langle \Delta_{Q_m,U_m}^2 \rangle_{\Delta k}} \Delta_{Q}^2 = r \, \Delta_{Q}^2
\end{equation}
where $\langle \, \rangle_{\Delta k}$ indicates the average over the $\Delta k$ range for both Stokes parameters. Similarly, $\Delta_{U}^2$ is constrained to $\Delta_{U'}^2$ at the corresponding frequencies.  In order to be consistent with the data, the simulated power spectra need to be scaled down by, at least, $r \sim 0.1$ at 126~MHz (Figure~\ref{fig:SphericallPS-David}; left panel), whereas they are already consistent (i.e. fainter) with the measurements at 164~MHz (Figure~\ref{fig:SphericallPS-David}; right panel).

These results can be used to improve our assumptions on the point source polarization fraction, allowing it to evolve with frequency. Defining $\gamma_{126}$ the polarization fraction at 126~MHz and recalling that
\begin{equation}
\Delta_{Q,U}^2 \, \propto \, \langle \gamma^2 \rangle
\end{equation}
and 
\begin{equation}
\langle \gamma^2 \rangle = \frac{2b}{3} \langle \gamma \rangle
\end{equation}
if $\gamma$ follows a uniform distribution between 0 and $b$, the comparison with the \cite{Moore2015} power spectra yields $\langle \gamma_{126} \rangle \le 0.1\%$, approximately a factor of three smaller than our model assumption.
It is interesting to note that such constraint qualitatively meets the expectations of Faraday depolarization models \citep[][]{Burn1966,Tribble1991} that predict the polarization fraction to decrease at longer wavelengths. Although the estimated leaked power spectra may still remain above the expected 21-cm power spectra, they are now more than one order of magnitude fainter than the initial model predictions.

One caveat of our comparison with real data is related to the role of ionospheric Faraday rotation. \cite{Moore2015} already pointed out that averaging visibilities over many days of observations to form polarized power spectra leads to significant depolarization due to time variable ionospheric Faraday rotation. Without any correction, polarized power spectra measured in an actual observation \citep[e.g.][]{Moore2015} can still be used to predict the leakage as we showed above, although they cannot be used to model the intrinsic sky properties and, therefore, straightforwardly predict the leakage contamination in a different 21-cm observation. In this respect, the constraints we placed on the average polarization fraction of extragalactic radio sources should be seen as constraints on the {\it effective} (i.e. modulated by ionospheric Faraday rotation) rather than the intrinsic fraction.

The effect of ionospheric Faraday rotation could be directly included in our formalism in equation~\ref{eq:intfresp} but we leave this for future work \citep{Aguirre2016}.

\section{Discussion and Conclusions}
\label{sec:conclusion}

We have presented a formalism to extend the delay--spectrum, visibility--based power spectrum estimator to full polarization, including the effect of polarized foreground leakage due to widefield primary beams. We applied our formalism to simulate power spectra from PAPER--like observations. We first used a total intensity source catalogue, predicting polarized power spectra in general agreement with observations in \cite{Kohn2016}. We then modeled polarized (Galactic and extragalactic) foregrounds using recent low frequency observations and predicted the corresponding power spectrum leakage, particularly focusing on the contamination for a 30~m baseline. 

We found that an ``EoR window" can be defined in terms of polarization leakage from diffuse Galactic foreground as its contamination falls quickly below $\sim 1$~(mK)$^2$ at $k > 0.3 \, h$~Mpc$^{-1}$, i.e. significantly below the fiducial range of 21-cm models. The existence of such EoR window is due to the fact that current observations find significant diffuse polarization only at low Faraday depths, i.e. $\phi \lesssim 12$~rad~m$^{-2}$ corresponding to $k_\parallel \lesssim 0.06 \, h$~Mpc$^{-1}$. Bright, diffuse emission found at high Faraday depth values would appear at proportionally higher $k_\parallel$ modes, narrowing (or jeopardizing) the EoR window. Current deep observations, however, set the presence of polarized diffuse emission to be below $\sim 0.1$~K at $\phi > 5$~rad~m$^{-2}$ \citep{Jelic2015}, supporting our model assumptions. 

In the case of point source leakage, an EoR window cannot be identified because point sources show emission essentially at any $k_\parallel$ value due to their broad RM distribution, making polarized point sources a potentially more serious contamination than diffuse emission. %- opposite to what has been historically considered. 
The magnitude of such leakage depends, however, significantly on the average point source polarization fraction for which only upper limits are currently available in the $100-200$~MHz range. By treating such upper limits as actual measurements, our model predicts a worse case scenario where point source polarization leakage is higher than the contamination due to Galactic emission at any $k$ mode for a 30~m baseline.

The comparison with polarized power spectra from \cite{Moore2015} constrains the observed (i.e. uncorrected for ionospheric Faraday depolarization) average polarization fraction at $\nu = 126$~MHz to be $\langle \gamma \rangle < 0.1\%$, leaving upper limits to the 21-cm leakage that are between one and two orders of magnitude greater than the 21-cm signal in the $7.7 < z < 10.3$ range.

Our current simulations do not include the depolarization effect due to ionospheric Faraday rotation average over multiple nights of observations, therefore all our predictions should be regarded as worst cases in terms of contamination to the 21-cm power spectrum.

Finally, our work provides a tool to predict the level of leakage expected in actual 21-cm observations by {\it forward modeling} the polarized foreground emission through the instrument model \citep[see][for relevant examples]{Pindor2011,Bernardi2011,Sullivan2012}: in the case of polarized point sources, for example, the observed average polarization fraction needs to be known in order to predict the leakage. We indicate three ways to determine the observed average polarization fraction:
\begin{itemize}
\item by best fitting the predicted polarized power spectra to the polarized power spectra measured in actual observations as we showed here with the \cite{Moore2015} data;
\item by imaging the polarized sky without correcting for ionospheric Faraday rotation: this directly provides a measurement of the average point source polarization fraction;
\item by applying an ionospheric Faraday rotation model to a polarized point source model realization whose average polarization fraction is provided by independent observations. 
\end{itemize}
Although future predictions might require further corrections to this first order picture, our model offers a way to account for polarization leakage in 21-cm power spectrum observations to be applied to future observations with PAPER, the Hydrogen Epoch of Reionization Array \citep{DeBoer2016} and, potentially, the Square Kilometre Array \citep{Koopmans2015}.

\section*{Acknowledgements}
We thank an anonymous referee for helpful comments that improved the manuscript. The authors also thank Tobia Carozzi for his help with the FEKO simulations and Adrian Liu for comments on the manuscript. CDN is supported by the SKA SA scholarship program. GB acknowledges support from the Royal Society and the Newton Fund under grant NA150184. This work is based on the research supported in part by the National Research Foundation of South Africa (Grant Numbers 103424). This research was supported by NSF grant AST-1440343. This research is supported by the South African Research Chairs Initiative of the Department of Science and Technology and National Research Foundation.

\bibliographystyle{apj}
\nocite{*}
\bibliography{Bibliography}

\end{document}